# Optimizing parameter search for community detection in time evolving networks of complex systems

**Italo'Ivo Lima Dias Pinto[1], Javier Omar Garcia[1,*], Kanika Bansal[1,2,*]**

[1] US DEVCOM Army Research Laboratory, Aberdeen Proving Ground, MD, USA.
[2] Computer Science and Electrical Engineering, University of Maryland, Baltimore County, MD, USA.

*Authors to whom correspondence should be addressed: javier.o.garcia.civ@army.mil, phy.kanika@gmail.com.

## Abstract

Network representations have been effectively employed to analyze complex systems across various areas and applications, leading to the development of network science as a core tool to study systems with multiple components and complex interactions. There is a growing interest in understanding the temporal dynamics of complex networks to decode the underlying dynamic processes through the temporal changes in network structure. Community detection algorithms, which are specialized clustering algorithms, have been instrumental in studying these temporal changes. They work by grouping nodes into communities based on the structure and intensity of network connections over time aiming to maximize modularity of the network partition. However, the performance of these algorithms is highly influenced by the selection of resolution parameters of the modularity function used, which dictate the scale of the represented network, both in size of communities and the temporal resolution of dynamic structure. The selection of these parameters has often been subjective and heavily reliant on the characteristics of the data used to create the network structure. Here, we introduce a method to objectively determine the values of the resolution parameters based on the elements of self-organization. We propose two key approaches: (1) minimization of the biases in spatial scale network characterization and (2) maximization of temporal scale-freeness. We demonstrate the effectiveness of these approaches using benchmark network structures as well as real-world datasets. To implement our method, we also provide an automated parameter selection software package that can be applied to a wide range of complex systems.

## Introduction

Coordination of foundational elements within a complex network is essential for a variety of systems from biological to social systems to environmental ecosystems (Boccaletti et al., 2014). Importantly, these networks are rarely ever static in nature; instead, they evolve over time, adapting to changing environmental constraints and contextual dynamics. The nature of this temporal evolution can be described with several techniques including percolation methods (Adamcsek et al., 2006), which characterize the transition from being fragmented to densely connected in a network; event-based methods (Asur et al., 2009), which characterize sporadic bursts in network reorganization; and dynamic community detection, which distills complex adjacency matrices into dynamic communities. The measured statistical dependency between elements of a network across time can represent information sharing, resource (re)allocation, trait similarity, amongst others, enabling the system as a whole to adapt and change in response to external forces. This ability to evolve and adapt is crucial for the long-term success and sustainability of many complex systems, as it allows it to overcome challenges and remain relevant in a constantly changing world (Baffy & Loscalzo, 2014). The manner in which a network changes over time provides an insight into how the system operates and how it can be improved and optimized for maximum efficiency and effectiveness or dissolution and degradation (Garcia et al., 2018).

Dynamic community detection, first introduced by Mucha et al. (2010), is a generalized approach to investigate temporal changes in complex networks. It has been used to describe a variety of network evolutions including task dependent brain networks (Bassett, Wymbs, et al., 2013) and social network dynamics (Gilbert et al., 2011). In neuroscience, it has been used to characterize several cognitive phenomena including memory (Braun et al., 2015), learning (Bassett et al., 2011), psychopathology (Li et al., 2022), chronic behavior change (Cooper et al., 2019) and most recently, online belief transformation (Lima Dias Pinto et al., 2022).

One of the primary benefits of this approach is that it harnesses an assumption of modularity across temporal and spatial scales. Previous studies have found that modular organization, as opposed to homogeneously connected networks, allows for rapid change and is crucial for rapid adaptation, robustness to perturbations, and efficient information processing (Hinne et al., 2015; Kashtan et al., 2007; Kashtan & Alon, 2005). Such organization is a typical feature of the human brain which evolved under pressures for adaptability, energy efficiency, and cost minimization, and it is also foundational to many other complex systems (Ghavasieh et al., 2020). However, dynamic community detection remains somewhat restricted to the field of network neuroscience and has not been fully utilized in other fields or contexts. The primary limitation in the deployment of this method is the requirement of a parameter search where the structural and temporal parameters of the system must be decided (Garcia et al., 2018). Here, we aim to overcome this limitation by providing objective criteria to determine these parameters.

In our solution, we harness the idea of self-organized criticality which is another property of many complex systems, apart from modularity, where large composite systems of many elements display emergent behavior that more holistically describes the dynamical system (Bak & Chen, 1991). Previous research has shown that large and complex systems display this behavior from geological explorations of earthquake dynamics (Sornette & Sornette, 1989) to neural firing and cognition (Shew & Plenz, 2013), to even describing the weather (Andrade et al., 1998). Here, we provide an analytically-sound conceptual framework for investigating such principles of complex systems, bridging the algorithms of dynamic community detection with elements from self-organized criticality.

In this paper, we first provide an overview of a form of clustering based on modularity maximization and its implementation using the Louvain algorithm (see Figure 1D; Blondel et al., 2008) and also the generalization that enables the investigation of dynamic community structures. Next, we explore properties of how a complex network reconfigures, including the *scale-freeness* of temporal processes, to propose a method that finds the optimal temporal and spatial scales in which to deploy dynamic community detection. We next validate this method on synthetic dynamic community structures used as benchmarks. Finally, we use the dynamic community detection algorithm combined with our optimization method on four datasets, two from neural recordings using two methods that vary in temporal and spatial resolution, a multi-person behavioral task, and an ant-communication network. These test cases span a wide range of "information exchange" between nodes of the network and vary in complexity of the signal, temporal dynamics, and spatial resolution.

# I. Detection of communities in a time evolving network using modularity maximization

To estimate the partitioning or communities within a network, Newman and Girvan (Newman & Girvan, 2004) proposed a modularity function (*Q*) that measures the extent or tendency of a network to separate into modules that have a high connectivity within and low connectivity between. This function was further generalized by Reichardt and Bornholdt (Reichardt & Bornholdt, 2006) to include a resolution parameter such that:

$$Q = \frac{1}{2m} \sum_{ij} \left[ A_{ij} - \gamma \frac{k_i \, k_j}{2m} \right] \delta(C_i, C_j), \qquad (1)$$

where, $A_{ij}$ is the weighted edge between nodes $i$ and $j$, $k_i$ is the degree of node $i$, $C_i$ is the community affiliation of node $i$, $m$ denotes the sum of the edge weights of the network, and $\delta$ denotes the Kronecker delta, which equals 1 if i = j, and 0 otherwise. Here, $\gamma$ is the resolution parameter which was introduced to overcome the resolution limit of the original modularity function (with $\gamma = 1$). In general, it allows for the tuning of the community sizes detected such that lower values of $\gamma$ yield larger communities and higher values yield smaller communities. A wide range of literature relies on maximizing this function to extract communities or partitions of a network (Garcia et al., 2018; Newman, 2006) by aiming to obtain an optimal partitioning such that the estimated modularity is higher than that of a *null model*. Often, a null model is created from a random network with the same distribution of node degrees as the original network (Newman, 2006), disrupting the modular properties of the network but retaining the general "connectedness" of the network.

Mucha et al. (Mucha et al., 2010) extended this function for time-evolving networks to perform dynamic community detection such that the modularity function changes to:

$$Q_{multilayer} = \frac{1}{2\mu} \sum_{ijlr} \left[ \left( A_{ijl} - \gamma \frac{k_{il} k_{jl}}{2m_l} \right) \delta_{lr} + \delta_{ij} \omega \right] \delta(C_{il}, C_{jr}), \qquad (2)$$

where, the indices $l$ and $r$ denote consecutive time layers. In the above expression, $m_l$ is the sum of the edge weights of layer $l$ and $\mu$ is the sum of the edge weights of all time layers.

Note that this equation (2) has an additional parameter $\omega$ along with the $\gamma$ parameter from equation (1). These resolution parameters play an important role during the extraction of communities from dynamic networks and put resolution limits on how small segments can be detected as separate communities. The resolution parameter $\gamma$ is often termed as the *structural* resolution parameter and it sets a resolution limit for the communities detected within a single layer of the network. Whereas $\omega$ is a *temporal* resolution parameter representing the weights of the between layer edges (or temporal edges) that connect a node with itself between consecutive network layers (Fig. 1B).

## (A) Critical role of the resolution parameters

While the real-world networks demonstrate modular architecture, extracting the partitioning of networks is not straightforward and, in fact, different partitioning schemes can be detected with different choices of parameters. For example, the role of the resolution parameters is depicted in Fig. 1C. As mentioned, $\gamma$ allows for the tuning of the community sizes within a layer, where smaller values of this parameter bias the communities to be larger in size (i.e., more nodes) and lower in count. Therefore, a value too low would detect a single community composed of all nodes within the network and a value too high would place every node in its own community (see Fig. 1C). Similarly, $\omega$ tunes how consistently a node is part of the same community across consecutive layers, where a higher value produces communities that would last for longer time periods (i.e., layers), while a lower value would make the network layers relatively independent leading to communities disappearing quickly (see Fig. 1C). In combination, these parameters determine the overall structure of the communities; if $\gamma$ is too high or too low, no meaningful structure would emerge within layers and, similarly, if $\omega$ is too high or low, no meaningful structure would emerge across layers.

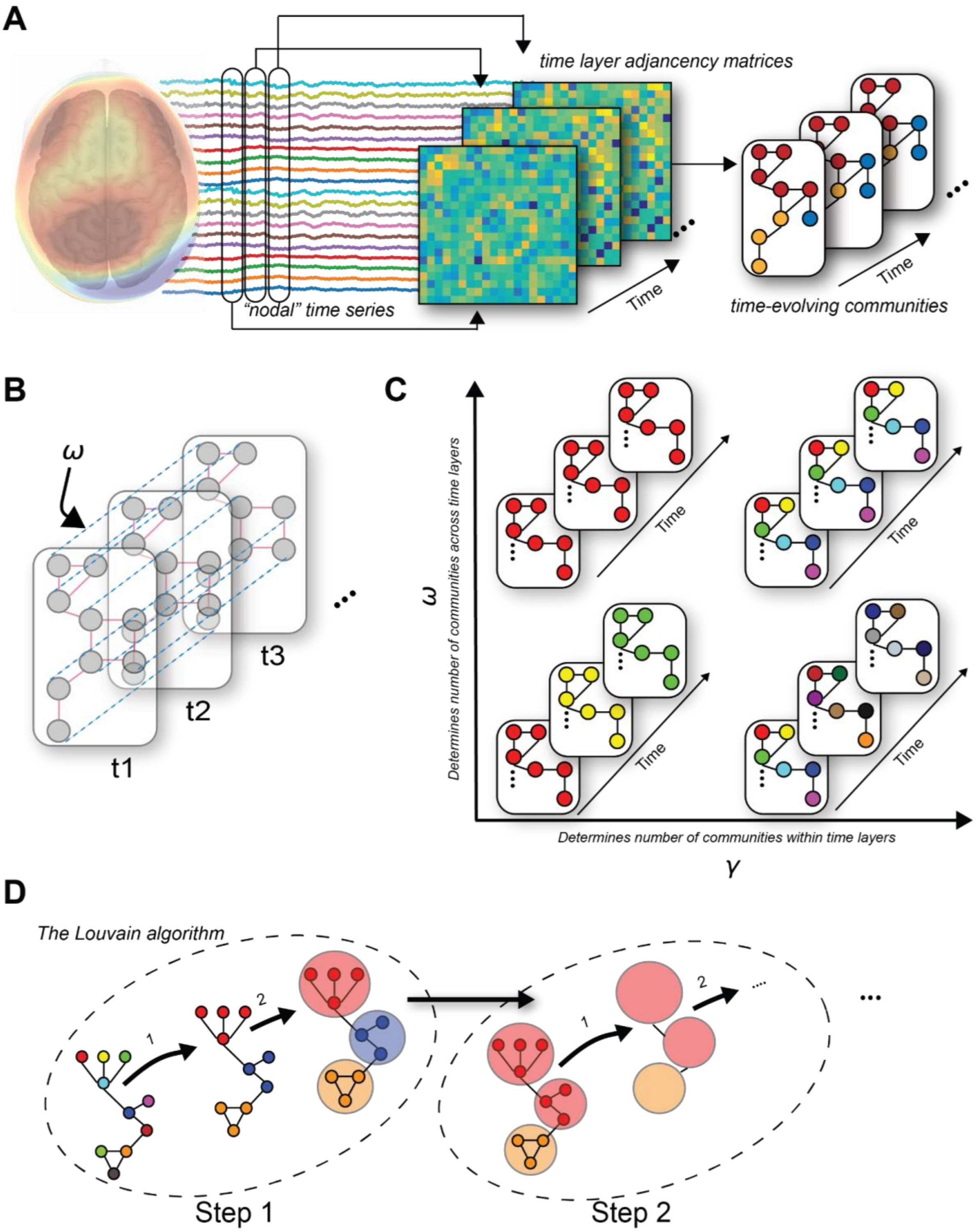

**Figure 1:** A typical dynamic community detection pipeline. (A) First, the data, for example, time series data representing activity from different regions of the brain, is segmented into time windows or layers and adjacency/connectivity matrices that represent statistical relationship between network nodes are calculated for each time layer. These adjacency matrices are used to estimate the dynamic community structure. (B) Across adjacent layers, a node is connected with itself through a link with strength $\omega$. This parameter $\omega$ acts as a temporal resolution parameter in addition to the structural resolution parameter $\gamma$, as described in the text. (C) Both $\omega$ and $\gamma$ play a crucial role in determining the dynamic community structure of the system. The structural resolution parameter $\gamma$ determines the size of the communities within a layer. Low (high) values of $\gamma$ produce communities that are large (small) in size and low (high) in number (see x-axis). On the other hand, the temporal resolution parameter $\omega$ determines how temporally constrained or independent the network nodes are. High values of $\omega$ make network nodes temporally constrained leading to communities that span large number of temporal layers, whereas, low values of $\omega$ make network nodes temporally independent leading to short lived communities across temporal layers. (D) One of the very popular algorithms to estimate community structure is the Louvain algorithm which uses two steps in every iteration as depicted here. Starting with every node in its own community, in step 1, nodes are combined in communities with other nodes such that these combinations increase the modularity of the network. When no further increase in modularity is possible step 2 starts. During step 2 a new network structure is determined such that the nodes in the same community are treated as a single node and all the links between two communities are summed to represent the link between the *new defined nodes*. Steps 1 and 2 are repeated for the new community structure until no modularity increases above a user defined tolerance level is possible.

## (B) Lack of objectivity in assigning resolution parameter values

While these resolution parameters play an important role in governing the community structure outcome, thus far, in the literature, their choice has been subjective or context dependent. For example, in network neuroscience, a variety of cognitive processes have been described using rapid reconfigurations in network structure and dynamic community detection has been used as a tool to capture these reconfigurations. Various metrics have been defined to summarize the outcome of the dynamic community detection; for example, flexibility describes how flexibly the nodes change their community allegiance over time and it has been shown to relate to various cognitive processes (Braun et al., 2015). Clearly, the choice of resolution parameters could impact the flexibility of network nodes, however, these choices have been optimized for robust detection across iterations or between conditions (Bassett, Porter, et al., 2013), which could overlook temporal-spatial scales related to other neural computations. In some cases, a value of 1 is assigned to these parameters (Telesford et al., 2016) and in other cases, a large parameter search is launched on the subset of the data and values of these parameters are chosen such that there are not too many or too few communities after modularity maximization (Garcia, Ashourvan, et al., 2020). In this paper, we propose an automated optimization method and provide a package to implement it in Python to objectively choose the values of these parameters. Importantly, our proposed methodology reduces biases for any specific temporal and/or spatial scale during the estimation of communities that might come from subjective selection of the resolution parameters. We will further elaborate on this point in the coming sections.

## (C) Dynamic community detection using generalized Louvain algorithm

One of the most popular and successful community detection algorithms is the so-called Louvain algorithm (Huang et al., 2021), first introduced by Blondel et al (Blondel et al., 2008), it received its name from the authors' affiliation. The Louvain algorithm is defined as a greedy modularity maximization algorithm, in which the communities merge until no further increase in modularity is possible. As portrayed in Fig. 1D, starting with every node in its own community, each iteration of the Louvain algorithm is composed of two steps: in the first step, nodes change their community affiliation to their neighbor's community affiliation in a way that leads to the largest increase of modularity possible for the neighbors of the current node. This process is repeated for each node until no further increase in modularity is possible. In the second step of the algorithm, communities are treated as a single node with link strength defined as the sum of link strengths of the nodes in the community. After the nodes are merged and the new network is defined, another iteration of the algorithm starts and the first step is applied to this new network. These interactions are repeated until the improvement in modularity goes below a predefined tolerance level (typically $10^{-10}$).

Here, we use this algorithm to demonstrate the parameter optimization and provide an automated dynamic community detection package using the modularity functions defined in Eq (2) based on a random network null model. Although we applied our parameter estimation method to the generalized Louvain algorithm, it can be easily adapted for other community detection algorithms or modularity functions as well.

# II. Optimization of the resolution parameters

We have discussed that the choice of resolution parameters generates scale biases in the community structure; low or high values of the structural resolution parameter generate larger or smaller communities within a layer while low or high values of the temporal resolution parameter generates shorter or longer communities across layers. We propose that the values of the parameters must be chosen to minimize these biases as much as possible. In other words, values of these parameters should be chosen in a way which generates no biases for a particular structural or temporal scale allowing for *scale-free* characteristics. In fact, scale-free architecture has been found to be a fundamental property of many real-world networks (Barabási & Bonabeau, 2003) and it is intuitive to look for scale-freeness in the properties of emergent community structure.

## (A) Optimization of the spatial resolution parameter ($\gamma$)

In Figures 2A-B, using brain network dynamics extracted from the human electroencephalography (EEG) data as an example (Lima Dias Pinto et al., 2022), we describe how changing the values of the structural resolution parameter affects the distribution of the size of the communities. Here, we performed dynamic community detection and assessed sizes of different communities that emerged across all the layers as a function of $\gamma$ by keeping $\omega$ = 1. As lower or higher values of $\gamma$ bias the community sizes to be larger or smaller, we observed that the distributions of community sizes were skewed to the right or left with a positive or negative skewness. We propose to minimize the scale bias by choosing the value of $\gamma$ that minimizes the absolute value of the estimated skewness.

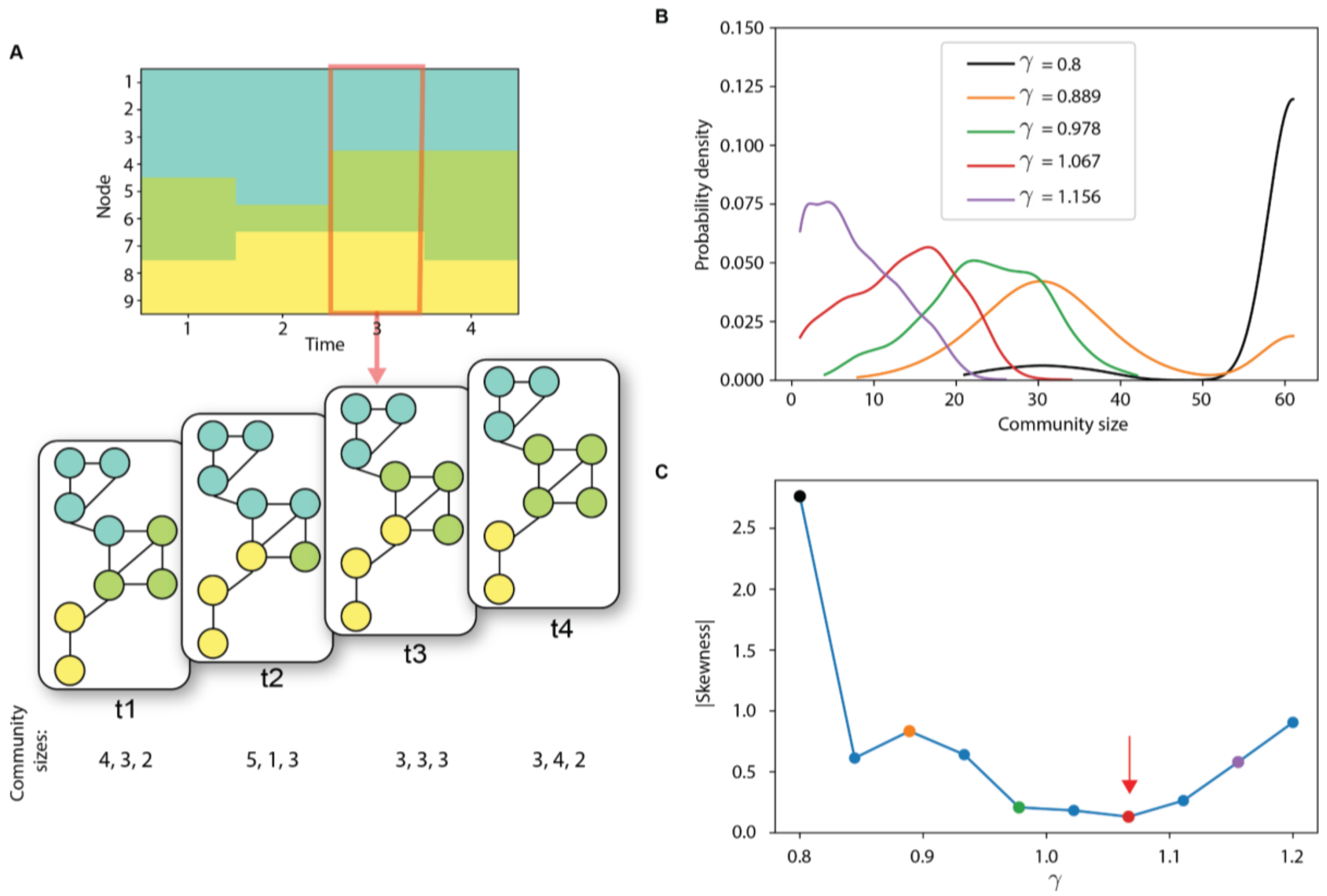

**Figure 2:** Optimization of the structural resolution parameter $\gamma$. (A) Toy example of community structure where each color represents a different community for a sample dynamic network composed of nine nodes and four time points. Graphical representation of the network is also shown with explicit community sizes within each temporal layer (bottom). (B) Probability density of community size as a function of the $\gamma$ parameter for brain network dynamics extracted from the human electroencephalography (EEG). Changes in the structural resolution parameter lead to changes in the community size distribution, with positive skewness for lower $\gamma$ values and negative skewness for higher $\gamma$ values. (C) Absolute values of the estimated *skewness* as a function of the $\gamma$ parameters. To minimize these trends towards large or small communities we search for a $\gamma$ value that minimizes the absolute value of the skewness (red arrow).

## (B) Optimization of the temporal resolution parameter ($\omega$)

In Fig. 3 we elaborate on the effect of $\omega$ on community structure using the same sample data as in Fig. 2. First, we define the *prevalence time* as the number of time layers a node spends in a given community before it changes its community affiliation (Fig. 3A). Larger values of $\omega$ produce communities that last longer while smaller values make communities short lived, therefore, probability distribution of prevalence time would have biases for relatively larger and smaller time scales as well.

We propose that the value of $\omega$ should be chosen to minimize any bias towards a specific time scale. Therefore, we propose to optimize $\omega$ such that the probability distribution of prevalence time is closest to a power law. Power-law distributions are used to describe scale-free characteristics of the measured quantity (Clauset et al., 2009) and have been found to describe many real-world processes like neurological and networks including social and technological (Clauset et al., 2009). A power-law probability distribution is given by:

$$P(x) = C\,x^{-\alpha x}, \quad (3)$$

where, $\alpha$ is a non-negative quantity which represents the exponent of the power-law.

To conduct this parameter search and estimate the closeness to a power-law, we first fit the probability distribution to a power-law distribution (Figures 3B-C). Subsequently, we generate a synthetic set of probabilities from an ideal power-law equation with the same prevalence time data and the exponent determined by the fitted value. We then evaluate the divergence between the actual and synthetic distributions by calculating the mean of the distances between the cumulative distribution functions of the actual (empirical) and synthetic distributions. We call this quantity mean residual and an example of this is shown in Fig. 3C. We choose the value of $\omega$ that produces the smallest mean residual.

## (C) Automation of the parameter search

Our Python-based package identifies the optimal parameters through a two-step, iterative refinement process. Initially, the community size skewness is calculated for a series of evenly spaced points within a user-defined $\gamma$ range (with $\omega$ = 1), and the algorithm selects the $\gamma$ value with the lowest skewness. Utilizing this estimated $\gamma$ value, the scale-freeness of the prevalence time is calculated, and the algorithm searches for the $\omega$ value that best conforms to a power-law, as previously detailed. To further refine the search, additional iterations are conducted using the points immediately higher and lower than the established $\gamma$ and $\omega$ values as the new boundaries for the search range, ultimately allowing the algorithm to pinpoint more accurate values for both parameters up to a user defined tolerance.

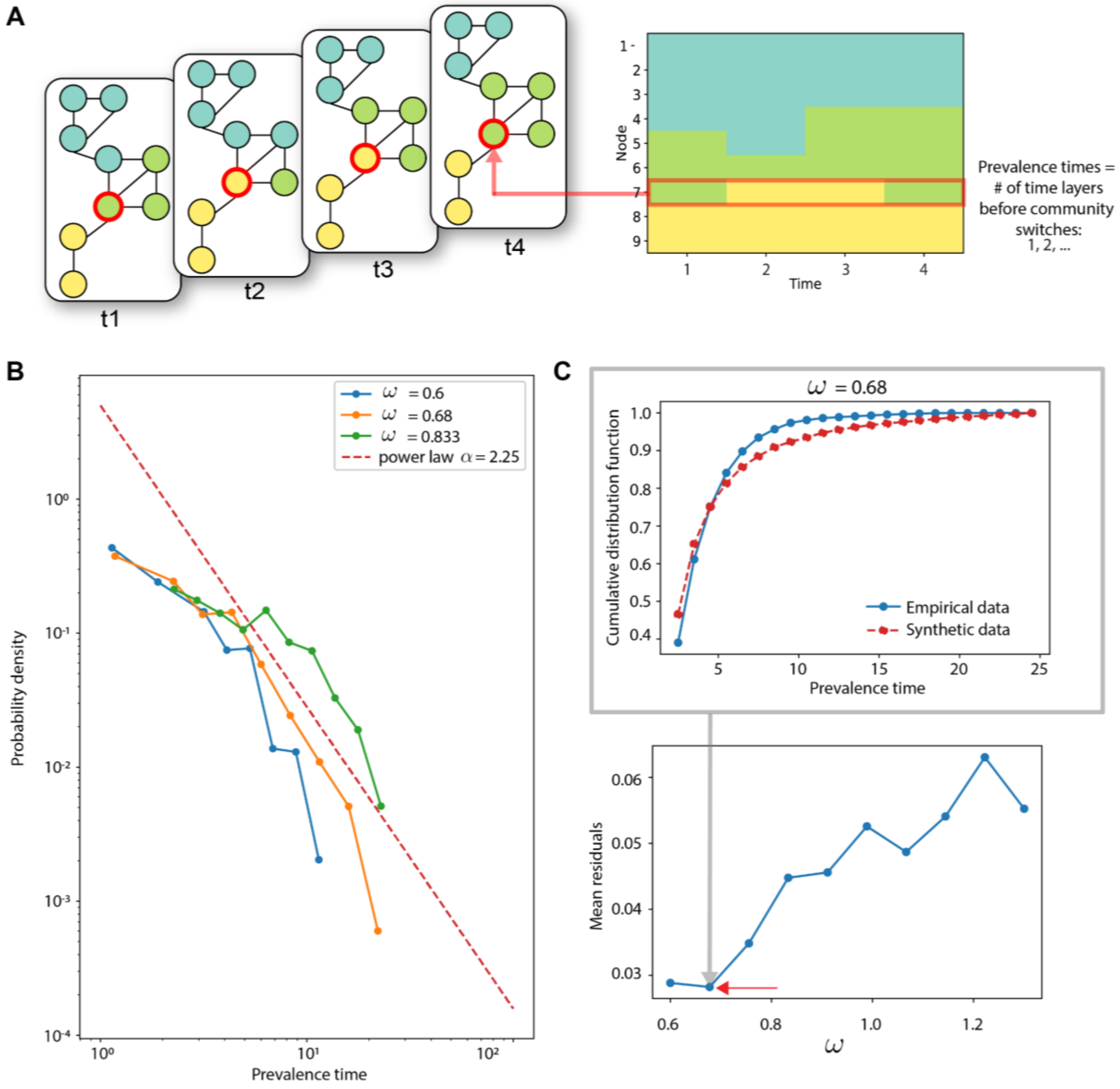

**Figure 3:** Optimization of the temporal resolution parameter. (A) Visual depiction of changing communities across time. Community change events are defined as events in which a node changes its community assignment. We define prevalence time as the number of layers between consecutive community change events for a node. (B) Probability density as a function of the prevalence time for different values of $\omega$. We optimize $\omega$ by aiming for a distribution that best matches a scale-free distribution (power-law) depicted as a red-dotted line. To implement this, we first fit a power-law to the prevalence time probability distribution and then generate a synthetic probability distribution by using the ideal power-law with the same prevalence time data and exponent from the fit. (C) To estimate the closeness of the empirical (actual) distributions of prevalence time to a power-law, we estimate the mean residuals as a function of $\omega$. Mean residual is the mean of the difference between cumulative distribution function for actual and synthetic data (top). We choose the value of $\omega$ that produces the lowest mean residual. This is an approximate measure of the "scale-freeness".

# III. Validation of the proposed methodology: comparison with benchmarks

In order to test and validate our parameter estimation method, we used temporal benchmark network models (Granell et al., 2015). These benchmark networks were specifically designed to test typical dynamic changes that can occur in evolving networks. The models are: (1) Grow-Shrink with some communities growing in size while other communities shrinking (Fig. 4A); (2) Merge-Split with communities merging in a single community and then splitting again (Fig. 4B); and (3) Mixed model which is a mix of the first two models (Fig. 4C). We performed the parameter detection on the models proposed in (Granell et al., 2015) using $q = 4$ communities with $n = 80$ total nodes and probability of links inside communities $p_{in} = 0.5$ and outside communities $p_{out} = 0.05$.

Figures 4A-C depict the benchmark communities or the ground truth using three different models as described above. In Figures 4D-F we show the communities we obtained using our parameter optimization method. A visual comparison of the community structures between the first two rows of Fig. 4 demonstrates that the obtained communities are close to the model communities in most scenarios, providing validation of our method. To further assess the quality of the obtained communities, we use the variance of information as described below.

## (A) Comparison with the benchmark models

The variance of information is an information-theoretic metric used to estimate the similarity (or dissimilarity) between two partitions. Here, we mathematically describe the variance of information between two partitions S and S' by using the concepts of partition entropy $H(\mathcal{S})$ and mutual information $I(\mathcal{S},\mathcal{S}')$ as depicted in Fig. 4G.
First, in the case of community labels, the partition entropy is defined as:

$$H(\mathcal{S}) = -\sum_{k=1}^{K} P(k) log\, P(k), \tag{4}$$

where, k represents a given community and P(k) represents the probability of a node to be in community k. Mutual information is defined as:

$$I(\mathcal{S},\mathcal{S}') = \sum_{k=1}^{K}\sum_{k'=1}^{K'} P(k,k') log \frac{P(k,k')}{P(k)P'(k')}\ , \tag{5}$$

where, P(k,k') represents the joint probability such that a node in S is in community k and the same node in S' is in community k'. The variance of information (VI) is defined as:

$$VI(\mathcal{S},\mathcal{S}') = H(\mathcal{S}) + H(\mathcal{S}') - 2I(\mathcal{S},\mathcal{S}'). \tag{6}$$

*VI* can be viewed as a measure of distance between two partitions, such that if the correspondence of the partitions is exact, the variance of information between them is 0, for more details of the properties of this metric we refer to (Meilă, 2007), which defines the variance of information as a metric to measure distances between different clusterings.

In Figures 4H-J we show the variance of information as a function of resolution parameters. The variance of information was calculated between the actual and obtained structures for each temporal layer and then we calculated the temporal mean of the variance of information (Meilă, 2007) to assess the quality of the partition obtained for different choices of the resolution parameters. Figures 4H-J show low values of the mean variance of information for the optimized resolution parameters we obtain, validating our methodology.

Please note that in case of merge-split model (Figures 4B, 4E), when the community structure undergoes an abrupt transformation involving substructures that rapidly merge and split within a single time point, the algorithm underperforms. This suboptimal performance can likely be attributed to the unique and unrealistic size and time prevalence of the communities in this model, which significantly deviate from our assumptions of minimal skewness and power-law distributed dynamics.

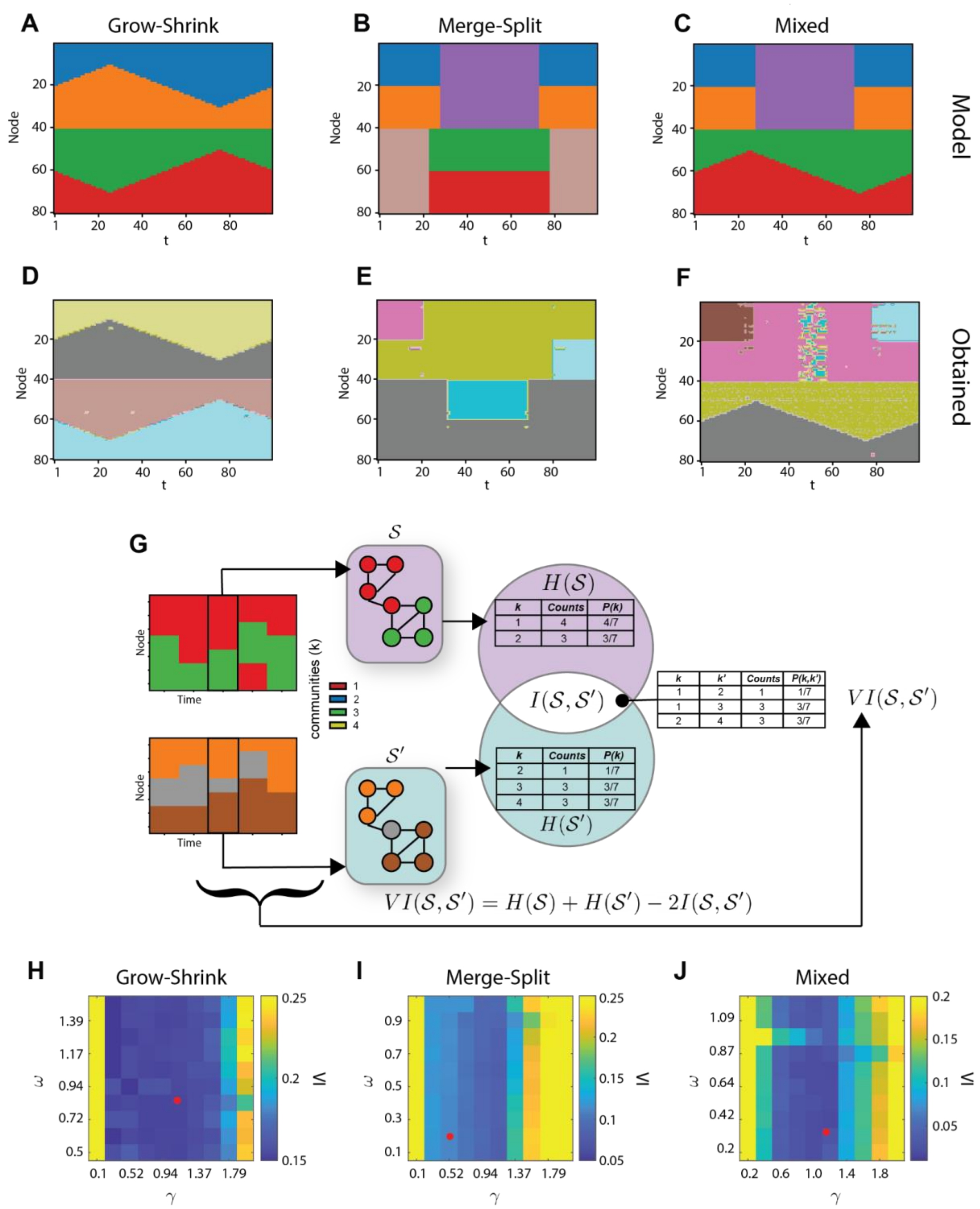

**Figure 4:** Validating our methodology using benchmark models. (A)-(C) Benchmark model networks with distinct community structures. Here, (A) represents the Grow-Shrink model with some communities growing in size while other communities shrinking; (B) represents Merge-Split model with communities merging in a single community and then splitting again; and (C) represents mixed model which is a mix of the first two models. (D)-(F) Obtained community structures for each model using generalized Louvain algorithm along with our parameter optimization method. A visual comparison with model networks suggests agreement between model and obtained community structures. (G) To assess the quality of obtained community structure, we estimate mean variance of information between the model

and obtained community structures. A schematic on how to calculate the variance of information (VI) is depicted here (Meilă, 2007). $S$ and $S'$ represent different partitions to be compared and H(S) represents the partition entropy for a given partition S. Different communities are represented by k and P(k) represents the probability of a node to be in community k. P(k,k') represents the joint probability distribution such that a node of S is in community k and the same node in S' is in community k'*. I(S,S') represents the mutual information between S and S'. (H)-(J) Mean variance of information, i.e., mean of VI calculated across all the temporal layers, as a function of $\gamma$ and $\omega$. Red dots represent the optimized values chosen by our algorithm which coincide with low values of mean variance of information.

# IV. Application of the proposed methodology to real data examples

The central theme of this deployment of dynamic community detection is in integrating elements from network science while anchoring the parameter search with a conceptual framework borrowed from complexity science. We chose four different datasets to span several elements that have been shown to display (i) network changes that dynamically reconfigure in a system due to many combined processes (i.e., neural underpinnings of human cognition), (ii) emergent properties for group success (i.e., team-based information gathering and behavior), and (iii) adaptive communication (i.e., ant communication networks). To understand the applicability of our technique, in the following, we apply it to these real-world examples an discuss out findings.

## (A) Application to the human neuroimaging data to uncover brain dynamics

The brain is often considered the most intelligent known complex system. One of the properties that often underlies this complexity is that it can rapidly reconfigure its dynamics to perform everyday tasks that are critical to its primary purpose, so-called *intelligence*. Two common techniques to measure human brain functioning are the blood oxygen level dependent (BOLD) response within the brain during functional magnetic resonance imaging (fMRI) as an indirect measure of neural activity and the electroencephalographic (EEG) measurements that can capture the direct population-level responses of the brain's network dynamics at the scalp. EEG has very high temporal resolution but very low spatial resolution (EEG), while BOLD has excellent spatial but poor temporal resolution. These two are complementing methods for non-invasive human neuroimaging. Both of these methods are commonly used in their respective cases to describe a variety of neural behavior associated with behavioral change (Braun et al., 2015; Cooper et al., 2019; Garcia, Ashourvan, et al., 2020; Garcia, Battelli, et al., 2020; Lima Dias Pinto et al., 2022). Critically, it has previously been shown that dynamic community reconfigurations extracted from both methods have displayed characteristics relevant to cognition (e.g., Zhang et al., 2016) and even have predictive qualities for behavioral outcome that follows (e.g., Lima Dias Pinto et al., 2022).

Here, we applied our method to a dataset containing concurrent EEG and fMRI recordings from an individual. EEG data was recorded from 61 sensors at a sampling frequency of 640

Hz which we used after a commonly used preprocessing pipeline to generate time evolving networks in the form of weighted connectivity matrices. Network connectivity was estimated by using the weighted phase locking index (wPLI) (Vinck et al., 2011) over 10 seconds non-overlapping time windows for a total duration of 49 minutes. In case of the fMRI data, the brain was parcellated into 200 regions or nodes derived from the Schaefer parcellation (Braun et al., 2018) and 14 subcortical regions. The BOLD signal from these 214 regions were used to obtain time evolving network layers by using wavelet coherence (Müller et al., 2004) between every two regions for 10 seconds non-overlapping time windows.

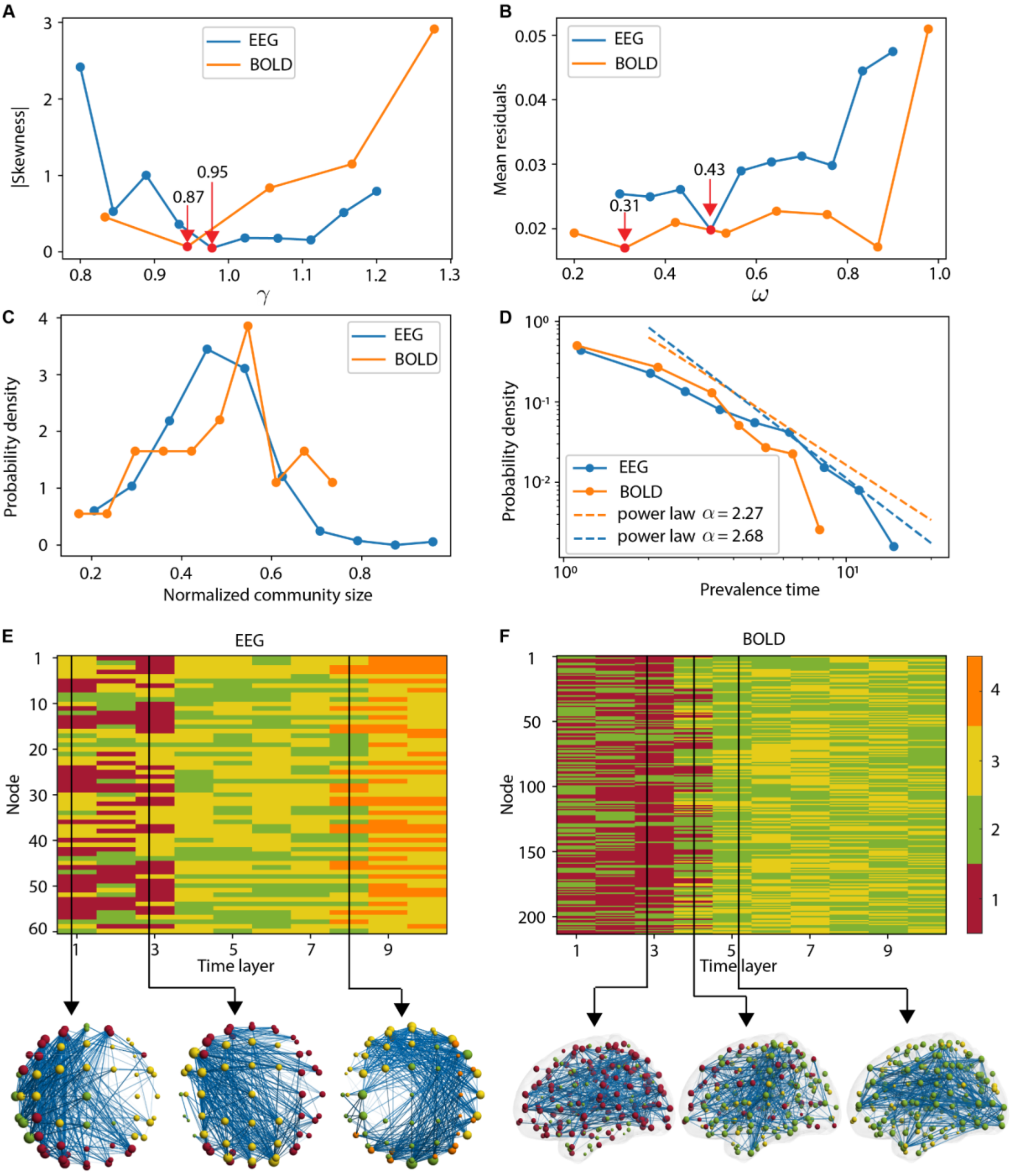

**Figure 5:** Dynamic community detection applied to the multimodal neuroimaging dataset with optimization of parameters. (A) Absolute value of the community size distributions skewness as a function of $\gamma$ for EEG as well as BOLD data; indicating the minima in red. (B) Mean residuals representing the deviation from scale-freeness values as a function of $\omega$; indicating the minima found in red. (C) The distribution of community sizes for the optimized $\gamma$ value of 0.95 (EEG) and 0.87 (BOLD). (D) Distribution of the prevalence time of nodes for the optimized $\omega$ value of 0.43 (EEG) and 0.31 (BOLD) that fits to power law distributions with $\alpha$ = 2.68 and 2.27 respectively. (E)-(F) Community structure for each modality, indicating the localization of each node of the network for the first 100 sec of EEG and f-MRI data (10 windows). Below is depicted the spatial distribution of top 5% of connections at 3 different time windows to visualize the dynamic changes in community demarcation and associated connectivity patterns.

From the temporal structure of adjacency matrices obtained with the described procedure, we ran the parameter search and found $\gamma = 0.87$, $\omega = 0.31$ for the BOLD data and $\gamma = 0.95$, $\omega = 0.43$ for the EEG data (Figures 5A-B). The distributions of community size normalized by the total number of nodes, for the optimized values of $\gamma$ are shown in Fig. 5C and the distributions of prevalence time for the optimized values of $\omega$ are shown in Fig. 5D along with the ideal power-law distributions. The community structure for each modality is depicted in Figures 5(E)-(F) for the first 100 seconds with different colors indicating different communities. Spatial distribution of these communities within the brain can be seen using the brain plots at three different instances. These plots highlight how different regions of the brain communicate dynamically making widespread communities.

## (B) Application to the ants contact network to uncover the structure of their colonies

Our second example refers to a contact network of a colony of ants during a 41 days experiment (Mersch et al., 2013). Ant colonies are one of the most intriguing complex systems. They demonstrate self-organization and the process of this emergent behavior has piqued the interest of many complex systems researchers. Mersch et al., used a video tracking system that tracked how the ants make contact with each other and with this contact network, using infomap community detection, they discovered that the colony had two robust communities which are made of *nursing* and *forager* ants. Further visual analysis of the data allowed them to discover a third community of *cleaner* ants which interact more often within the community while also maintaining consistent communication with the other two communities. They also reported that the ants tended to change roles and migrate between communities.

We believe that the information regarding these communities and role change, as encoded in how the ants communicate, can be easily and robustly extracted with our dynamic community detection approach. We used the data in Mersch et al. to track the dynamics of the colony. The network was constructed such that each node represents an ant and each time layer a day of the experiment, an entry of the adjacency matrix is 1 if that pair of ants touched each other during that day and 0 otherwise. Even though our approach can be used with binary networks, for consistency with other datasets and to reduce noise, we generated weighted networks by using a rolling window of 5 days by connecting ants with a weighted edge which represents the mean of if or not the ants touched each other for 5 days. Results are discussed in Fig. 6. Figures 6A-B show the absolute value of skewness as a function of $\gamma$ and mean residuals as a function of $\omega$. Optimized values picked by our algorithm are highlighted with red arrows with $\gamma$ = 1.01 and $\omega$ = 0.26. Probability Distribution of community sizes and prevalence time for these optimized values are shown in Figures 6C-D.  In Fig. 6E we show how the community structure changes with time (days).

We observe that at any given time, there exist predominantly three communities. Aligned with the findings from Mersch et al., these communities could represent nursing, forager, and cleaner ants. Importantly, we see how these communities change over time, likely due to the change in their roles. While our findings mimic the findings from the Mersch et al., our analysis is relatively direct showing three robust communities instead of two. Additionally, during this experiment some ants died before the last recorded day and an interesting feature emerges from the use of the generalized Louvain algorithm, i.e., the dead ants are represented as a continuous line in the community structure of Fig. 6E as they maintain affiliation with their last community. This effect is due to such nodes being disconnected to all other nodes within the network having only connections with themselves in the consecutive time layers.

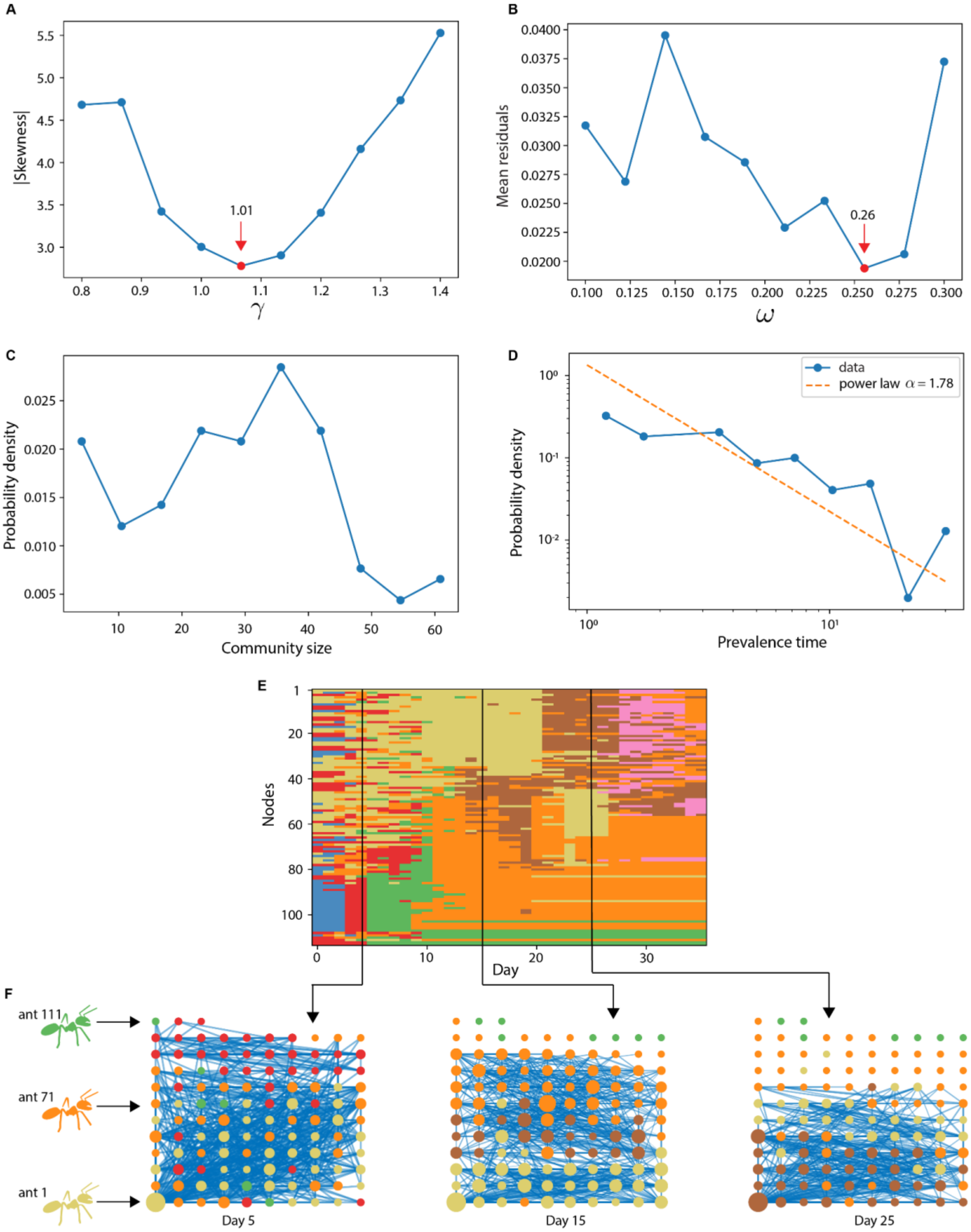

**Figure 6:** Dynamic community detection applied to the contact networks of ants with optimization of parameters. (A) Absolute value of the community size distribution skewness as a function of $\gamma$; indicating the minima in red. (B) Mean residuals representing the deviation from scale-freeness values as a function of $\omega$; indicating the minima found in red. (C) The distribution of community sizes for the optimized $\gamma$ value of 1.01. (D) Distribution of the prevalence time of nodes for the optimized $\omega$ value of 0.26 that fits to a power law distribution with $\alpha$ = 1.78. (E) Community structure of the ant contact network across days. At any given day, predominantly three communities can be identified. Importantly, the structure of these communities changes with time, likely indicating how the roles of the ants change within the colony. (F) Partial snapshots of the network for a few ants on three different days, indicating how the network connectivity and hence the community structure evolves.

## (C) Application to the social game experiment

Our next example is a social game data set (Bai et al., 2019) where the subjects play a game called Resistance (Indie Boards & Cards) in which the players must interact and use social cues to determine high level social constructs such as deception. Critical to this gameplay is to understand the intention of other players which has been studied by monitoring eye movements and eye contact between individuals (Bai et al., 2019). In this case, the adjacency matrices were obtained through eye-tracking devices worn by each player. This dataset allows us to inspect many interacting complex systems (individuals) that are limited in the exchange of information in general.

For each individual, the eye tracking device recorded which other players they were looking at in 1/3 of a second time window and we obtained an adjacency matrix using this information. In the original form, the adjacency matrices are asymmetric, representing who was looking and who was being looked at. To account for a pairwise interaction, we constructed a symmetric weighted adjacency matrix $A$ such that between individual nodes $i$ and $j$ $A_{ij} = A_{ji} = 0.5\ (A'_{ij} + A'_{ji})$ where $A'_{ij}$ and $A'_{ji}$ represent the elements of the asymmetric adjacency matrix.

Results for the dynamic community detection for this data are shown in Figure 7. Fig. 7A displays the absolute values for skewness as we change $\gamma$ and Fig. 7B shows the value of the mean residual as we change $\omega$. We found the optimized value for $\gamma$ to be 0.785 however, for $\omega$, the algorithm picked a very low value of $5.48 \times 10^{-11}$. This value was picked to represent a minimum in the mean residual as a function of $\omega$ and while the values of the mean residual are comparable to other datasets, this low value highlights a weaker connection of a node to itself across temporal layers likely due to the changing individual behavior during the game. Figures 7C-D show the distributions of community size and prevalence time of nodes for the optimized parameter values. Fig. 7E shows the community structure evolution for the first few time layers (rounds) of the time series. Formation of smaller groups can be clearly visualized representing how individuals are visually attending to each other. This kind of analysis can be used to assess how a team of individuals operates and how subgroups emerge, opinions are formed, and consensus is built.

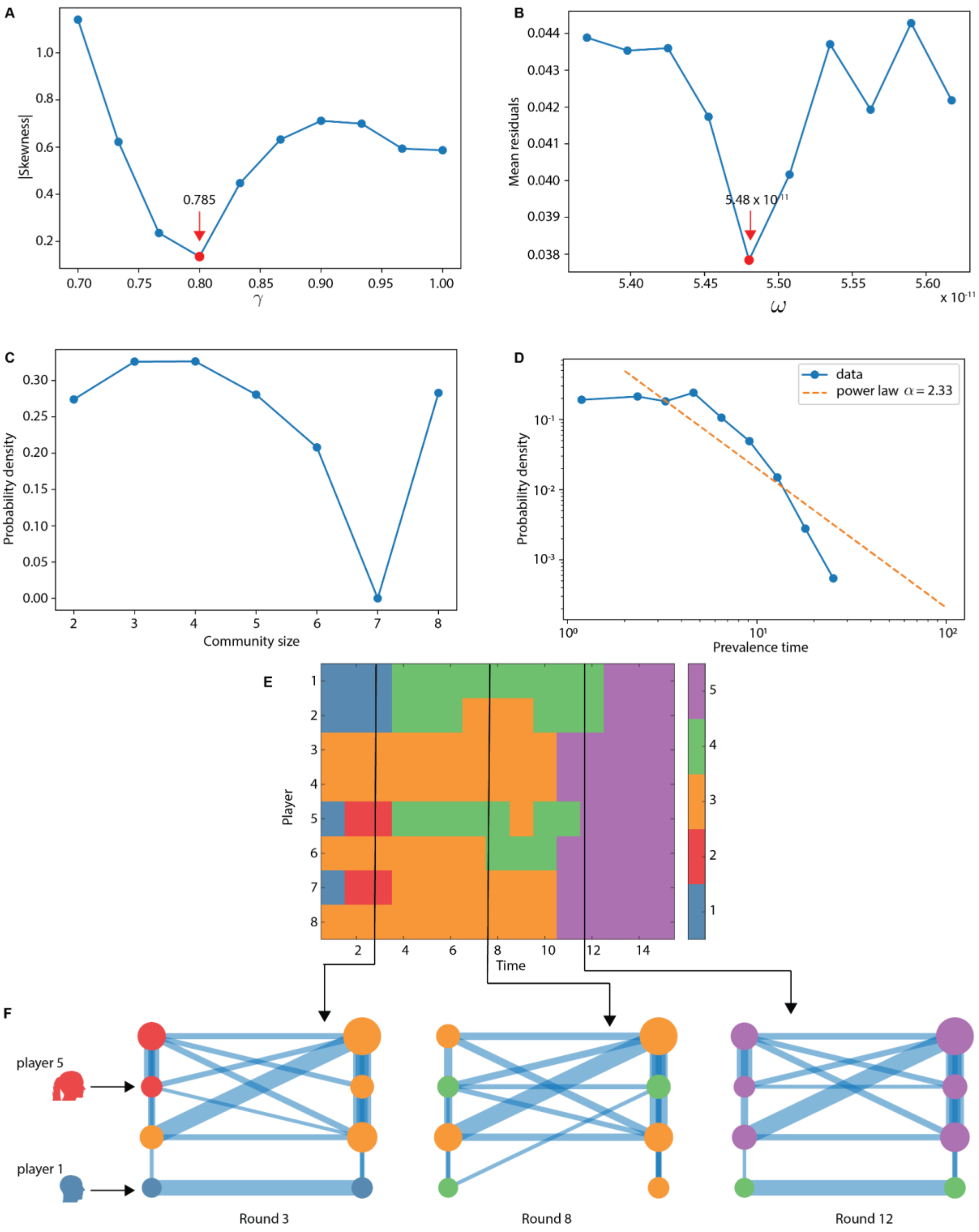

**Figure 7:** Dynamic community detection applied to the human eye contact networks during a social game play. (A) Absolute value of the community size distribution skewness as a function of $\gamma$; indicating the minima in red. (B) Mean residuals representing the deviation from scale-freeness values as a function of $\omega$; indicating the minima found in red. (C) The distribution of community sizes for the optimized $\gamma$ value of 0.785. (D) Distribution of the prevalence time of nodes for the optimized $\omega$ value of $5.48 \times 10^{-11}$. (E) Community structure of the eye contact network. (F) Partial snapshots of the network for a few players on three different instances indicating how the network connectivity and hence the community structure evolves.

## V. Discussion and conclusions

We proposed an objective framework to optimize parameter search for dynamic community detection that aims to reduce any scale biases and attempts to optimize to near-scale-free dynamics. This computational choice is not arbitrary, but in fact, is inspired from many real-world networks and processes, including neurological processes, that show scale-free characteristics (Andrade et al., 1998; Bak & Chen, 1991; Shew & Plenz, 2013; Sornette & Sornette, 1989). Using the proposed framework, we provide a Python based package for automated implementation of the parameter search and dynamic community detection which also allows for individual inputs.

We validated and demonstrated the application of our framework using a variety of models and sample data. We particularly showed how the dynamic community detection method, combined with our objective parameter optimization, provides interesting insights into the dynamics of different systems that operate at different spatio-temporal scales. While the reconfiguration of brain networks has been studied using dynamic community detection in multiple papers (Bassett et al., 2011; Braun et al., 2018; Cooper et al., 2019; Garcia, Ashourvan, et al., 2020; Garcia, Battelli, et al., 2020; Lima Dias Pinto et al., 2022; Telesford et al., 2016), this method has not been employed in other fields of network science as frequently. We propose that combined with our automated optimization of parameters, dynamic community detection method can be insightful in a variety of contexts. and it provides a unifying framework in which various complex systems can be analyzed and compared.

The objective nature of the proposed method also provides new opportunities for comparing dynamics across different complex systems. By automating the selection of resolution parameters and optimizing scale-freeness, the method provides a standardized approach that can be applied consistently across various systems. This consistency is crucial when making comparisons, as it ensures that any observed differences in dynamics are due to inherent properties of the systems themselves, rather than differences in methodology or measurement device. The objective parameter selection process could also provide insights into the scale differences between complex systems. If the optimal parameters for two systems or its states are significantly different, this could suggest that the systems operate on different scales. For example, one process might exhibit dynamics that change rapidly over time, requiring a low temporal resolution parameter, while another process might have more stable dynamics that change slowly, requiring a high temporal resolution parameter. Similarly, differences in the optimal structural resolution parameter could indicate differences in the size or complexity of the communities within the systems. Critically, due to the objective nature of the parameter search, impacts of the spatiotemporal resolution of the measurement may be minimized along with the dependencies on other computational choices such as the temporal width and gap between temporal layers.

In addition, the proposed method could potentially be used to identify relationships or correlations between different complex systems. For example, if two systems consistently require similar resolution parameters, this could suggest that they share similar dynamic properties or underlying structures. This could be particularly useful in fields like ecology or systems biology, where researchers are often interested in understanding the relationships between different ecosystems or biological networks (Girvan & Newman, 2002; Mersch et al., 2013).

However, it's important to note that while the objective nature of the method facilitates comparisons between systems, these comparisons still need to be interpreted with caution. Complex systems are often influenced by a wide range of factors, and it's possible that differences in dynamics could be due to external influences rather than inherent properties of the systems. Therefore, any conclusions drawn from these comparisons should be validated through further research or experimentation. Moreover, we used the generalized Louvain algorithm to implement the detection of communities with our example data. While this algorithm is one of the most common, it is a greedy algorithm and may affect the ultimate choice of parameters. Previous research has deployed different algorithms for the detection of communities, each of which brings different limitations and benefits to its estimated community structure (Huang et al., 2021); we believe that parameter values that are optimized for a particular algorithm might not be generalizable to a different algorithm.

In conclusion, our proposed method offers a promising, robust approach to investigating dynamic complex systems. However, a careful consideration of the resolution parameters is essential which requires an understanding of the system's structural and temporal scales and their potential interactions with the measurement process. These interactions may vary, moreover, the measurement process could only represent a partial view of the complex system in question. Future work could focus on developing more efficient parameter search algorithms and exploring the applicability of this method to a wider variety of complex systems.

## Acknowledgements

This research was sponsored by the US DEVCOM Army Research Laboratory and was completed under Cooperative Agreement Numbers W911NF2020067 (I.L.D.P.), and W911NF-17-2-0158 (K.B.). The views and conclusions contained in this document are those of the authors and should not be interpreted as representing the official policies, either expressed or implied, of the US DEVCOM Army Research Laboratory or the U.S. Government. The U.S. Government is authorized to reproduce and distribute reprints for Government purposes notwithstanding any copyright notation herein.

## Author declarations section

### Conflict of Interest

The authors have no conflicts to disclose.

### Author Contributions

**Italo'Ivo Lima Dias Pinto:** Conceptualization (equal), Data Curation, Formal Analysis, Methodology (equal), Software, Validation, Visualization (equal), Writing/Original Draft Preparation (equal), Writing/Review & Editing (equal). **Javier Omar Garcia:** Conceptualization (equal), Funding Acquisition (equal), Methodology (equal), Project Administration (equal), Resources, Supervision (equal), Visualization (equal), Writing/Original Draft Preparation (equal), Writing/Review & Editing (equal). **Kanika Bansal:** Conceptualization (equal), Funding Acquisition (equal), Methodology (equal), Project Administration (equal), Supervision (equal), Visualization (equal), Writing/Original Draft Preparation (equal), Writing/Review & Editing (equal).

## Code availability statement

The code, scripts, and algorithms used in this study are freely available for academic and non-commercial use. All codes used for data analysis and generation of results have been deposited in GitHub at the following repository: https://github.com/italoivo/DCD.

In the repository, you will find detailed comments and instructions for usage within the README file. For any inquiries, suggestions, or difficulties regarding the usage of the codes, please contact the corresponding author of this paper or raise an issue in the GitHub repository. While we strive to ensure the code is as user-friendly and well-documented as possible, we are available to provide further clarification or support as needed.

Please note that the use of the code should be acknowledged by citing this paper in any resulting publications.

Please understand that while the code is provided as is, we cannot guarantee it will function identically in all environments or be compatible with future versions of dependencies. Users are responsible for ensuring the correct environment setup.